\documentclass[
aps,%
11pt,%
final,%
notitlepage,%
oneside,%
twocolumn,%
nobibnotes,%
nofootinbib,%
superscriptaddress,%
noshowpacs,%
centertags]%
{revtex4}

\usepackage{multirow}
\renewcommand{\baselinestretch}{1.0}

\begin{document}

\selectlanguage{english}


\title{Ultra-Flat Galaxies Selected from RFGC Catalog. II. Orbital
Estimates of Halo Masses}

\author{\firstname{I.~D.}~\surname{Karachentsev}}
\email{ikar@sao.ru} \affiliation{\saoname}

\author{\firstname{V.~E.}~\surname{Karachentseva}}
\affiliation{Main Astronomical Observatory of National Academy of
Sciences of Ukraine, Kiev, 03680 Ukraine}

\author{\firstname{Yu.~N.}~\surname{Kudrya}}
\affiliation{Astronomical Observatory of Taras Shevchenko National
University of Kiev, Kiev,  04053 Ukraine}

\received{January 21, 2016}  \revised{February 25, 2016}

\onecolumngrid
{\scriptsize
ISSN 1990-3413, Astrophysical Bulletin, 2016, Vol. 71, No. 2, pp.129-138 @ Pleiades Publishing, Ltd., 2016\\
Original Russian Text @ I.D.Karachentsev, V.E.Karachentseva, Yu.N.Kudrya 2016,\\
 published in Astrofizicheskii Byulleten, 2016, Vol.71, No.2.}

\begin{abstract}
 We used the Revised Flat Galaxy Catalog (RFGC) to select 817 ultra-flat (UF)
edge-on disk galaxies with blue and red apparent axial ratios of
$(a/b)_B > 10.0$ and $(a/b)_R > 8.5$. The sample covering the whole
sky, except the Milky Way zone, contains 490 UF galaxies with
measured radial velocities. Our inspection of the neighboring
galaxies around them revealed only 30 companions with radial velocity
difference of $\mid\Delta V\mid<500$~km\,s$^{-1}$  inside the
projected separation of $R_p < 250$~kpc. Wherein, the wider area
around the UF galaxy within $R_p < 750$~kpc contains no other
neighbors brighter than the UF galaxy itself in the same velocity
span. The resulting sample  galaxies mostly belong  to the
morphological types Sc, Scd, Sd. They have a moderate rotation
velocity curve amplitude of about
  $120$~km~s$^{-1}$  and a moderate K-band luminosity of about
 $10^{10}L_{\odot}$. The median difference of radial velocities of
their companions is $87$~km~s$^{-1}$, yielding the median orbital
mass estimate of about $5\times10^{11}M_{\odot}$.
 Excluding six probable non-isolated
pairs, we obtained a typical halo-mass-to-stellar-mass   of UF
galaxies of about $30$, what is almost the same one as in the
principal spiral galaxies, like M\,31 and M\,81 in the nearest
groups. We also note that ultra-flat galaxies look two times less
``dusty'' than other spirals of the same luminosity.

\end{abstract}

\maketitle

\section{INTRODUCTION}

The population of thin (flat)  spiral galaxies is the most suitable
laboratory for the study of physical processes of the formation and
evolution of galactic disks. As noted by many
authors~\cite{kar1989:Karachentsev_n,kau2009:Karachentsev_n,sha2015:Karachentsev_n},
simple disks of galaxies with no visible signs of a bulge avoid the
regions with high environment density. An obvious reason for this is
a supposed
 lack in isolated disks of a noticeable
tidal perturbation from the nearest   neighbors, capable of ``warming
up'' the stellar disks  in the vertical direction. According
to~\cite{mat2000:Karachentsev_n,kar2002:Karachentsev_n,kau+2009:Karachentsev_n,kre2005:Karachentsev_n},
thin disks distinguish themselves among the other spiral galaxies by
decreased average surface brightness, blue color and low rotation
curve amplitudes. Spectroscopic observations of several ultra-thin
galaxies~\cite{goa1981:Karachentsev_n}
have shown that the effects of  emission line excitation by
large-scale shock waves are  mild in them.


Kormendy~\cite{kor2013:Karachentsev_n}
 has repeatedly stressed that the very existence of a large
population of massive galaxies, devoid of bulges, is a big problem
for the current theories of galaxy formation, where numerous mergers
of small objects  lead to a consistent growth of bulges. Comparing
the images of flat  galaxies obtained at the Hubble Space Telescope
with the images from the Sloan Digital Sky Survey, Sachdeva et
al.~\cite{sac2015:Karachentsev_n}
have found that over the past 8 billion years since the era of
$z\sim1$ until the present time, linear dimensions and masses of
galactic disks  have increased approximately twice. From there, the
authors have concluded that the dominant mode of growth in thin disks
is the accretion of intergalactic gas, rather than the process of
hierarchical merging of dwarf galaxies.


Detection of ultra-thin disks among the galaxies, oriented at
arbitrary angles to the line of sight is quite a challenge. The
surest way to do this is to use a sample of spirals seen edge-on. In
our previous paper~\cite{kara2016:Karachentsev_n}
we made a selection of the most flat galaxies, based on the Revised
Flat Galaxy
Catalog~\cite{kar1999:Karachentsev_n}. 
Among the 4236 RFGC objects, 817~ultra-flat$=$\,UF galaxies were
selected, whose apparent blue~($B$) and red~($R$) axial ratios
     satisfy the condition: \mbox{$(a/b)_B>10.0$} and \mbox{$(a/b)_R>8.5$}.
     This sample covers the entire northern and southern sky, except for the
region of the Milky Way \mbox{($|b|<10^{\circ})$}
 and has an  approximately $90$\% completeness up to the  angular diameter of
  \mbox{$a_B=1\farcm2$}~\cite{kara2016:Karachentsev_n}.
From the analysis of this fairly representative sample, we concluded
that about 60\% of UF galaxies have not got any  close neighbors
within the projection distance of
 $R_p=750$~kpc  and  the radial velocity differences of $\mid\Delta V\mid<500$~km\,s$^{-1}$.
 The rest of the UF sample (approximately $30$\%) is a part of scattered associations
 and filaments  along with the other brighter neighbors, and
only around $10$\% of UF galaxies are the dominant objects in the
dynamically bound multiple systems. We shall use the latter category
in the following to estimate the mass of the dark halo of ultra-flat
galaxies, which, to our knowledge, has never been previously
estimated.

\begin{figure}
 \onelinecaptionsfalse  \captionstyle{normal} \setcaptionmargin{5mm}
\includegraphics[width=0.95\columnwidth]{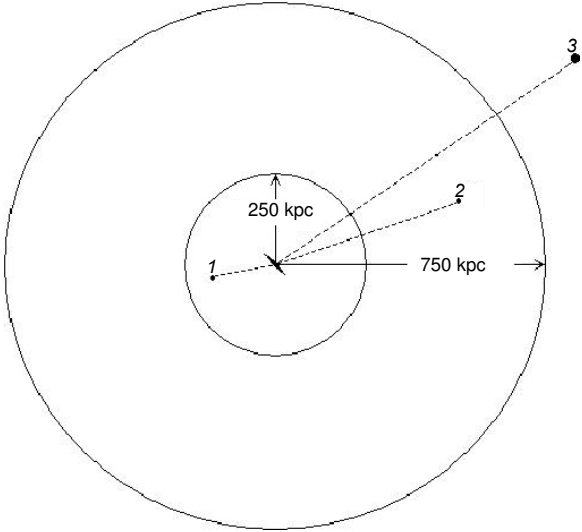}
\caption{The scheme, explaining the selection of physical satellites
around the ultra-flat galaxy.}
\end{figure}

\section{UF GALAXY ORBITAL MASS ESTIMATION}
The brightest spiral galaxies in the nearby groups, such as the
M\,31, M\,81, and NGC\,253 have a characteristic stellar mass of
\mbox{$M^*\sim 8\times10^{10}M_{\odot}$},  the typical  rotation
curve amplitude of \mbox{$V_m\sim 250$~km\,s$^{-1}$} and the halo
radius  of around
$250$~kpc~\cite{kar2014:Karachentsev_n,kar+kud2015:Karachentsev_n}.
These galaxies, which dynamically dominate their environment, have
quite noticeable bulges. The  halo-mass-to-stellar-mass ratio in them
is \mbox{$M_h/M^*\sim30$}~\cite{kar+kud2014:Karachentsev_n}.
By the $M^*$ and $V_m$  values,  ultra-flat spirals are significantly
inferior to the  ``host'' spiral galaxies in the nearby groups. This
gives us reason to believe that the characteristic halo radius around
the UF galaxies is less than $250$~kpc.


To search for the physical satellites related to the UF galaxies, we
used the following simple algorithm, illustrated by Fig.~1. The
considered ultra-flat galaxy should not have other  brighter galaxies
with the velocity difference of $\mid\Delta V\mid<500$~km\,s$^{-1}$
within the radius of $R_p=750$~kpc around it. Among the neighboring
fainter galaxies in the given range of radial velocities we
considered to be physically linked only the satellites which are
located within the projected separation of  $R_p=250$~kpc. Thus, in
the scheme of Fig.~1, galaxy \emph{1} is the  UF companion, and
galaxy \emph{2} may be either a UF companion,  or a companion of a
more massive galaxy \emph{3}. Of course, the criterium we used  can
not be called perfect. It may be met by a  ``UF + close neighbor in
the projection '' pair, the components of which are a part of a
scattered (non-virialized) association or a chain  of galaxies.

We conducted the search of satellites around 490 ultra-flat galaxies
with measured radial velocities  using the options of the NASA
Extragalactic Database (NED)({\tt www.ned.ipac.caltech.edu}).  The
linear projected separation of neighbors was determined under the
assumption that their radial distances are equal to the distance of
the UF galaxy,  $D_{\rm UF}=V_{\rm h}/H_0$ with the Hubble parameter
of \linebreak $H_0=73$~km\,s$^{-1}$\,Mpc$^{-1}$.
 The result of this massive search was the detection of only 30 suspected physical
satellites the data on which are  presented in Table~1. A small
 number of satellites once again evidences that  ultra-flat
galaxies without bulges are located in the regions of very low number
density of galaxies.

 The columns of Table~1 contain: (1)~the number
of the UF galaxy in the RFGC catalog; (2)~an abbreviated name of the
satellite galaxy or the name of the
 sky survey in which its radial velocity was measured; (3)~the heliocentric radial velocity of the UF galaxy
  and its satellite from the NED with the measurement errors~(km\,s$^{-1}$); (4)~the morphological type:
  for the RFGC galaxy---based on
the catalog data,  for the satellite according to our estimates; (5,
6)~the apparent axial ratios in blue and red bands according to the
RFGC; (7)~the apparent $B$-magnitude of the  RFGC galaxy from the NED
and our estimate of $B$-magnitude of the satellite; (8)~Galactic
extinction in the $B$-band; (9)~the apparent~$K_s$-band magnitude
from the 2MASS
survey~\cite{jar2000:Karachentsev_n,jar2003:Karachentsev_n}; in the
cases where the 2MASS data were not available,  to estimate it we
used the relationship between the  rotation curve amplitude, $V_m$
and the $K$-luminosity (see Section 4);

\renewcommand{\baselinestretch}{0.85}
\setcaptionwidth{\linewidth}%
\setcaptionmargin{0mm} %
\onelinecaptionstrue %
\captionstyle{normal}
\begin{longtable*}{lr|r@{$\,\pm\,$}l|c|c|c|c|c|c|c|c|c|c|c|c}
\caption{Ultra-flat galaxies with their orbital mass estimates}\\
\hline
\multicolumn{2}{l|}{RFGC}   &\multicolumn{2}{c|}{\multirow{2}{*}{$V_h\pm\sigma$}}&  \multirow{2}{*}{$T$}& \multirow{2}{*}{$(a/b)_B$}& \multirow{2}{*}{$(a/b)_R$} &  \multirow{2}{*}{$B$ } & \multirow{2}{*}{$A_B$} &\multirow{2}{*}{  $K_s$}  & \multirow{2}{*}{$m_{21}$} & \multirow{2}{*}{ $V_m$}&\multirow{2}{*}{ $R_p$}&\multirow{2}{*}{$\log(M_{\rm orb})$} &\multirow{2}{*}{ $\log M^*$ }& \multirow{2}{*}{$\log(\frac{M_{\rm orb}}{M^*})$}\\
\multicolumn{2}{r|}{Neighb} &\multicolumn{2}{c|}{}  &  & & &    & & & & & & &  &\\
\hline (1)&            (2)&    \multicolumn{2}{c|}{(3)}&   (4) &
(5)&(6)&(7)&(8)&(9)&(10)&(11)&(12)&(13)&(14)&(15)\\ \hline
\endfirsthead
\caption{Ultra-flat galaxies with their orbital mass estimates}\\
\hline
\multicolumn{2}{l|}{RFGC}   &\multicolumn{2}{c|}{\multirow{2}{*}{$V_h\pm\sigma$}}&  \multirow{2}{*}{$T$}& \multirow{2}{*}{$(a/b)_B$}& \multirow{2}{*}{$(a/b)_R$} &  \multirow{2}{*}{$B$ } & \multirow{2}{*}{$A_B$} &\multirow{2}{*}{  $K_s$}  & \multirow{2}{*}{$m_{21}$} & \multirow{2}{*}{ $V_m$}&\multirow{2}{*}{ $R_p$}&\multirow{2}{*}{$\log(M_{\rm orb})$} &\multirow{2}{*}{ $\log M^*$ }& \multirow{2}{*}{$\log(\frac{M_{\rm orb}}{M^*})$}\\
\multicolumn{2}{r|}{Neighb} &\multicolumn{2}{c|}{}  &  & & &    & & & & & & &  &\\
\hline
(1)&            (2)&  \multicolumn{2}{c|}{(3)} &   (4) & (5)&(6)&(7)&(8)&(9)&(10)&(11)&(12)&(13)&(14)&(15)\\ 
\hline
\endhead
\hline
\endfoot
\hline
\endlastfoot
\hline
  99&                            & 5339 & 6  & Sc &  11.2 &  10.2   &15.03 & 0.09& 10.19& 14.6& 190 &  10.99  &         &                \\
  \multicolumn{2}{r|}{2dF}       & 5306 & 89 & BCD&       &         & 18.2 &     &      &     &     & & 33&  10.64   &  --0.35\\
\hline
 124&       & 7170&12 & Sb & 10.0  &  8.7   &14.64& 0.07&  9.83& 16.5& 294 &11.36 & & \\
 \multicolumn{2}{r|}{2MASX}     & 7225&64 & BCD& & & 17.3 & & & & & & 68&  12.03  &   0.67\\
\hline
 166&       & 9448&53 & Sc &  14.1 & 11.1   &16.29& 0.07&     & &   &11.01 & & \\
 \multicolumn{2}{r|}{GALEX}     & 9500&89 & BCD&  & & 18.6 & & & & & &114&  11.56  &   0.55\\
\hline
 239&       & 7083&45 & Sc &  10.1 & 10.4   &16.0 & 0.13& 12.15&    &    & 10.44  & & \\
 \multicolumn{2}{r|}{SDSS}      & 6935& 5 & Im & & & 17.6 & & & & &  &94&  12.39  &   1.95 \\
\hline
 365&       & 5439&15 & Scd& 10.0  & 10.7   &16.5 & 0.10& 13.66& 16.1&  93   & 9.60 & & \\
 \multicolumn{2}{r|}{MCG-2-5-36} & 5456& 4 & Sm & & & 16.5 & & & &  & &175&  10.78  &   1.18 \\
\hline
 625&       & 4955&13 & Sd &  11.2 & 11.2   &16.3 & 0.33& 11.92& 16.1& 128  & 10.22 & &  \\
 \multicolumn{2}{r|}{SDSS}      & 5042&18 & Im & & & 18.4 & & & & & &167&  12.17   &   1.95\\
\hline
 627&       & 5279& 6 & Sd &  13.7 & 12.3   &16.5 & 0.42& 13.86& 16.3&  87  &  9.50 & & \\
 \multicolumn{2}{r|}{UGC\,2397}   & 5117& 9 & Sm & & & 16.9  & & & &  & & 100&  12.50   &   3.00 \\
\hline
 722&       & 1873& 2 & Sd &  11.5 &  8.6   &15.33& 0.15& 11.27& 14.2& 101 &  9.60 & &   \\
 \multicolumn{2}{r|}{MCG-3-9-37}& 1866& 5 & Sm & & & 16.5  & & & &  & & 39&   9.34  &  --0.26\\
\hline
1000&       & 4121&11 & Scd& 12.5  & 11.5   &15.3 & 0.47&  9.63& 14.1& 250   & 11.00 & & \\
\multicolumn{2}{r|}{MCG+13-5-3} & 4267&25 & dE & & & 16.0  & & & &  & & 216&  12.73  &   1.73\\
\hline
1236&       & 2475&14 & Sd &  11.2 & 11.2   &16.5 & 0.09& 14.07& 16.2&  84  & 8.80 & &  \\
 \multicolumn{2}{r|}{UGC\,3940}   & 2453& 6 & Im & & & 16.5  & & & &  & & 189&  11.03   &   2.23 \\
\hline
1462&       &  596& 6 & Sdm& 10.1  &  9.3   &15.0 & 0.11& 11.24& 14.0&  48  &  8.62 & & \\
 \multicolumn{2}{r|}{SDSS}      &  588&34 & Im & & & 17.2  & & & &  & & 48&   9.46  &   0.84\\
\hline
1522&       & 7698&31 & Sd &  11.2 & 11.8   &16.7 & 0.07& 13.58&    &   & 9.93 & & \\
 \multicolumn{2}{r|}{2MASX}     & 8082&17 & BCD& & & 16.8  & & & &  & & 90&  13.17  &   3.24\\
\hline
1567&       & 3219&10 & Sd &  14.1 &  9.8   &16.6 & 0.16& 12.77& 15.4&  84   & 9.44 & & \\
 \multicolumn{2}{r|}{SDSS}      & 3505& 1 & BCD& & & 18.1  & & & &  & &182&  13.24  &   3.80 \\
\hline
1716&       & 7905&33 & Scd& 13.1  & 10.2   &16.3 & 0.05& 12.09&    &    & 10.55  & & \\
\multicolumn{2}{r|}{ SDSS}      & 7933&25 & Im & & & 17.9  & & & &  &  &58&  10.73  &   0.18\\
\multicolumn{2}{r|}{MCG8-18-65} & 7956&28 & Sd & & & 17.0  &  & & & &  &65&  11.30   &   0.75\\
\hline
1744&       & 3150&34 & Sd &  10.2 &  9.0   &17.0 & 0.05& 13.10& 16.7&  78     &  9.34 & &   \\
\multicolumn{2}{r|}{SDSS}      & 3127&14 & BCD& & & 18.2  & & & &  & &157&  11.00  &   1.66\\
\hline
1782&       & 9722& 4 & Scd& 11.5  & 11.2   &16.2 & 0.05& 12.87& 16.8& 184      & 10.40 & &  \\
\multicolumn{2}{r|}{SDSS}      & 9674&28 & BCD& & & 17.6  & & & &  & &117&  11.49  &   1.09 \\
\hline
1880&       & 5612& 5 & Sd &  13.9 & 11.2   &15.75& 0.05& 12.11& 15.8& 132      & 10.25 & &    \\
\multicolumn{2}{r|}{SDSS}      & 5528&34 & Im & & & 18.1  & & & &  & &190&  12.20  &   1.95 \\
\hline
1925&       & 4162& 1 & Sd &  11.7 & 11.7   &16.1 & 0.10& 13.23& 15.7&  88       &  9.51 & &     \\
\multicolumn{2}{r|}{U\,6054Not.1} & 4270& 8 & Sm & & & 18.0  & & & &  &  &13&  11.23   &   1.72 \\
\hline
2111&       & 5254& 4 & Sd &  12.2 & 11.0   &16.1 & 0.10& 13.24& 15.8& 105 & 9.70 & &  \\
\multicolumn{2}{r|}{SDSS}      & 5146&10 & Sdm& & & 17.9  & & & &  &  &99&  12.11  &   2.41 \\
\hline
2210&       & 2452& 2 & Scd& 11.6  & 10.1   &14.9 & 0.11& 12.37& 13.9&  94   & 9.38 & &  \\
\multicolumn{2}{r|}{UGC\,7133}   & 2567& 7 & Sd & & & 14.9  & & & &  & &204&  12.51  &   3.13\\
\hline
2474&       & 1642& 3 & Sm &  15.5 & 14.0   &16.04& 0.38& 12.10& 16.1&  65  & 9.07 & &  \\
\multicolumn{2}{r|}{GALEX}     & 1450&45 & Im & & & 16.8  & & & &  & & 74&  12.50   &   3.43 \\
\multicolumn{2}{r|}{GALEX}     & 1509&45 & BCD& & & 16.9   & & & & & &199&  12.62   &   3.55\\
\hline
2546&       & 6724& 6 & Sc &  11.7 &  9.5   &15.9 & 0.08& 11.93&   &   & 10.46 & &  \\
\multicolumn{2}{r|}{SDSS}      & 6857&30 & Sm & & & 18.4   & & & & & &183&  12.58  &   2.12 \\
\hline
2819&       & 5832& 2 & Scd& 11.7  & 10.5   &15.58& 0.11& 11.50& 15.7& 144  & 10.52 & &  \\
\multicolumn{2}{r|}{SDSS}      & 5746&30 & Ir & & & 18.8   & & & & & &107&  11.97  &   1.45\\
\hline
3021&       & 1915& 3 & Sc &  10.6 &  8.5   &13.9 & 0.41& 10.6 & 14.5& 113  &  9.91 & &  \\
\multicolumn{2}{r|}{UGC\,9979}   & 1961& 3 & Im & & & 14.6   & & & & & &116&  11.46  &   1.55\\
\hline
3087&       & 5694& 3 & Sc &  11.2 & 10.2   &16.7 & 0.23& 13.60& 15.6&  99 & 9.68 & &  \\
\multicolumn{2}{r|}{SDSS}      & 5728&27 & BCD& & & 18.7   & & & & &  &68&  10.97   &   1.29 \\
\hline
3444&       & 7178&10 & Sc &  11.1 &  9.6   &16.2 & 0.50& 11.74& 14.4& 199  & 10.61 & & \\
\multicolumn{2}{r|}{2MASX}     & 7074&45 & Sb & & & 16.2   & & & & &  &94&  12.08  &   1.47\\
\hline
4081&       & 4839& 6 & Sc &  10.4 &  9.1   &14.53& 0.39& 10.38& 14.3& 236 & 10.84 & &  \\
\multicolumn{2}{r|}{RFGC\,4082}  & 4720& 3 & Sdm& & & 16.0   & & & & & &192&  12.50  &   1.66 \\
\hline
4091&       & 4961& 2 & Scd& 11.1  & 10.1   &15.4 & 0.20& 11.30& 14.8& 134 & 10.49 & &  \\
\multicolumn{2}{r|}{2MASX}     & 4739&235& Sm & & & 16.0  & & & &  &  &80&  12.68  &   2.19\\
\hline
\end{longtable*}%
\renewcommand{\baselinestretch}{1.0}

\noindent(10)~the H\,I-flux (Jy\,km\,s$^{-1}$), expressed in
magnitudes according to the HyperLeda\footnote{{\tt
http://leda-univ-lyon1.fr}} as\linebreak \mbox{$m_{21} = -2.5\log
F(\mathrm{H\,I})+17.4$}; (11)~the rotation curve amplitude from the
HyperLeda (km\,s$^{-1}$); (12)~the log of the ultra-flat galaxy
stellar mass (in~$M_{\odot}$);(13)~the projected separation of the
satellite in kpc; (14)~the log of the orbital mass
estimate~(in~$M_{\odot}$);  (15)~the log of the
orbital-mass-to-stellar-mass ratio.

To estimate the orbital mass, we used the expression
\begin{equation}
    M_{\rm orb}=(16/\pi G)\Delta V^2 R_p,  
\end{equation}
valid at a random orientation of satellite orbits   and the RMS orbit
eccentricity value of  $\langle
e^2\rangle=1/2$~\cite{kar1987:Karachentsev_n};
here $G$ is the gravitational constant, and $\Delta V$ is the radial
velocity difference of the satellite and the UF galaxy.


When calculating the   luminosity of a galaxy in the $K$-band
\begin{equation}
\begin{aligned}
     \log L_K & =0.4[5\log D_{\rm Mpc}-K  \\  
          & +28.28+E(B-V)/2.93]
\end{aligned}
\end{equation}
we accept the absolute $K$-magnitude of the Sun,
\mbox{$m^K_{\odot}=3.28$}~\cite{bin1998:Karachentsev_n}
and a correction for the Galactic extinction according
to~\cite{sch1998:Karachentsev_n}.


As we can see from Table~1, only two ultra-flat galaxies (RFGC\,1716
and 2474) have two satellites each within
 $R_p=250$~kpc, while in the remaining 26 UF galaxies the ``escort'' is presented by
just one satellite.

\begin{figure}
\onelinecaptionsfalse  \captionstyle{normal} \setcaptionmargin{5mm}
\includegraphics[width=0.95\columnwidth]{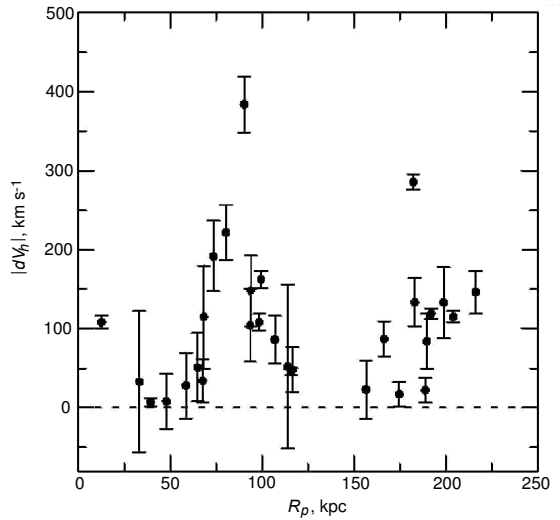}
\caption{The distribution of 30 satellites of ultra-flat galaxies by
the radial velocity difference modulus    and the projected
separation.}
\end{figure}

The distribution of ultra-flat galaxies and their satellites by the
projected separation and the radial velocity difference modulus is
shown in Fig.~2. The vertical bars on it correspond to the quadratic
sum of errors $\sigma_{\rm v}$ for the pair components. In some
cases, the error of the velocity difference is greater than the
difference modulus itself. As can be seen from Fig.~2, the radial
velocity differences are small. They exceed $250$~km\,s$^{-1}$ only
in two cases. The median value of the velocity difference modulus is
$87$~km\,s$^{-1}$. The orbital mass estimates for UF galaxies,
presented in column~(13) of Table~1 are characterized by a high
scatter, a considerable part of which is due to random projection
factors. The mean value of the orbital mass logarithm amounts to
$11.77\pm0.18$,
 which is close to the  value of the median logarithm,
$12.04$. As far as we know from the literature, these estimates are
the first estimates of the mass of the halo around the ultra-flat
galaxies made on the scale of their effective halo radius.

\section{MORPHOLOGICAL TYPES OF ULTRA-FLAT GALAXIES}

As shown by Heidmann et al.~\cite{hei1972:Karachentsev_n},
the maximum apparent axial ratio  $(a/b)_{\rm max}$  in spiral
galaxies increases along the Hubble sequence from the  Sa type to Sd
type, and then sharply drops  for the irregular structure spirals, Sm
type. According to~\cite{kud1994:Karachentsev_n},
the maximum apparent and the maximum intrinsic axial ratio  for
different types of spirals are characterized by the  following
values: $13.0$ and $14.1$ (Sb), $15.3$  and $17.2$ (Sbc), $19.2$  and
$22.0$ (Sc), $19.5$ and $22.4$ (Scd), $22.4$  and $27.0$ (Sd).


Among the 30 considered   ``UF galaxy + satellite'' pairs, in 26
cases an ultra-flat galaxy belongs to the Sc, Scd, Sd or $T = 5$,
$6$, $7$ morphological type on de~Vaucouleurs scale. With the average
error of classification of galaxies in the RFGC amounting to $\Delta
T = \pm1$~\cite{kar1997:Karachentsev_n},
we can assume that almost all the ultra-flat galaxies have a uniform
disk structure.

\begin{figure}
\onelinecaptionsfalse  \captionstyle{normal} \setcaptionmargin{5mm}
\includegraphics[width=0.95\columnwidth]{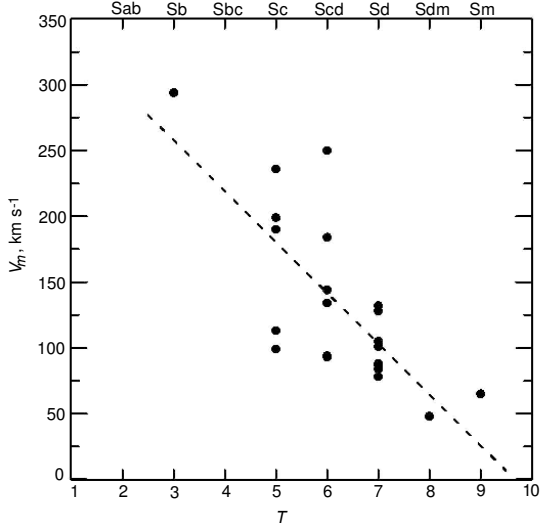}
\caption{The distribution of 23 UF galaxies from Table~1 by the
rotation curve amplitude and the morphological type.}
\end{figure}

Figure 3 shows the distribution of  UF galaxies from Table~1 by the
morphological type and the rotation curve amplitude   $V_m$. Despite
the small statistics, the figure shows a decrease of the rotation
curve amplitude from early to late types. For the spirals of the Sdm
and Sm types, the
  amplitude of  regular motion,   $V_m\sim 60$~km\,s$^{-1}$,
becomes comparable with the average velocity of turbulent motions,
$V_{\rm turb} \sim 15$~km\,s$^{-1}$, which impede the formation of
ultra-thin disks.


The morphological types of the satellites of UF galaxies in
column~(4) of Table~1 belong to even  later types than those of the
ultra-flat galaxies themselves. We have classified more than a half
of the satellites
 as Sm and BCD, what indicates that they dwell in the  phase of active
star formation.

\section{STELLAR MASSES OF ULTRA-FLAT GALAXIES}

To determine the total stellar mass of a galaxy, its $K_s$-band
luminosity is typically used, assuming that $M^*/L_K = 1.0
M_{\odot}/L_{\odot}$~\cite{bel2003:Karachentsev_n}. The 2MASX
catalog~\mbox{\cite{jar2000:Karachentsev_n,jar2003:Karachentsev_n}}
contains data on $K_s$-magnitudes for roughly $70$\% of UF galaxies.
However, being a shallow  sky survey, 2MASS underestimates the
luminosity of peripheric regions of galaxies, especially the blue
objects of low surface brightness, to which
 quite a few  UF galaxies belong. The top panel of Fig.~4 demonstrates
 the Tully-Fisher relation~\cite{tul1977:Karachentsev_n} between the \mbox{$K$-luminosity} according to the 2MASS and the logarithm
of the rotation curve amplitude for  23 galaxies from Table~1. The
regression line    for them is expressed as
\begin{equation}
\begin{aligned}
 \log(L_K/L_{\odot})&= 3.57(\pm0.21) \log V_m  \\  
              & + 2.55(\pm0.44),
\end{aligned}
\end{equation}
while the standard deviation from the regression line is $0.20$. The
parameters of this regression are close to the parameters obtained
from the more extensive sample of RFGC
galaxies~\cite{kar2002:Karachentsev_n}.

\begin{table*}
\setcaptionmargin{0mm} \onelinecaptionstrue \captionstyle{normal}
\caption{Ultra-flat galaxies with $K_s$-photometry by  Dalcanton  and
Bernstein}
\medskip
\begin{tabular}{r|r|l|c|c|r@{$\,\pm\,$}l|c|r@{$\,\pm\,$}l|c|c|c} \hline
 RFGC& \multicolumn{1}{c|}{$V_h$} &Type &$(a/b)_B$&$(a/b)_R$&\multicolumn{2}{c|}{$V_m\pm\sigma$} & $B_{27}$ & \multicolumn{2}{c|}{$K_{22}\pm\sigma$} & $E(B-V)$& $\log(L_K$)& $m_{21}$\\
\hline
\multicolumn{1}{c|}{(1)}&\multicolumn{1}{c|}{(2)}&\multicolumn{1}{c|}{(3)}&(4)&(5)&\multicolumn{2}{c|}{(6)}&(7)&\multicolumn{2}{c|}{(8)}&(9)&(10)&(11)\\
\hline
   73&  5287 & Sdm & 12.4 & 9.9  & 66&3   &17.52 & 15.42&0.27& 0.053 &  8.87   &16.94 \\
  267& 16186 & Sd  & 11.0 & 9.1  &  \multicolumn{2}{c|}{--}      &17.72 & 12.86&0.07& 0.049 & 10.86   &  --\\
  415& 11430 & Sd  & 11.4 &10.0  &146& 5  &17.20 & 13.33&0.16& 0.051 & 10.38   &16.69\\
  430&  5620 & Scd & 10.1 &11.0  &106& 4  &17.19 & 13.39&0.07& 0.112 &  9.74   &16.63  \\
  500&  4316 & Sd  & 11.2 & 8.9  & 84& 6  &17.94 & 14.10&0.12& 0.201 &  9.24   &16.08\\
  676&  7619 & Sd  & 15.1 &13.1  &  \multicolumn{2}{c|}{--}      &17.71 & 13.50&0.11& 0.126 &  9.96   &  --\\
  769&  6101 & Sd  & 13.9 &10.5  &160& 5  &16.15 & 11.36&0.14& 0.234 & 10.64   &14.97\\
 1587&  4329 & Sd  & 11.0 & 9.6  & 83& 3  &17.02 & 13.91&0.22& 0.038 &  9.30   &16.74\\
 1672&  2156 & Scd & 11.8 & 8.6  & 97& 3  &15.39 & 11.81&0.14& 0.057 &  9.54   &15.38\\
 1761&  3755 & Sd  & 17.1 &14.9  &131& 3  &15.33 & 11.41&0.12& 0.010 & 10.17   &14.89\\
 2260&  1598 & Sd  & 13.3 &11.9  & 81& 3  &15.16 & 12.04&0.37& 0.023 &  9.18   &15.54\\
 2295&  4240 & Sd  & 20.4 &15.9  &142& 5  &15.72 & 11.52&0.15& 0.020 & 10.23   &14.80\\
 2928&  2023 & Sd  & 14.8 &13.0  & 62& 2  &15.24 & 12.53&0.93& 0.051 &  9.19   &15.20\\
 3064& 10387 & Scd & 10.9 & 9.7  &228& 8  &16.64 & 12.54&0.28& 0.042 & 10.60   &16.04\\
 3274&  2781 & Sd  & 10.2 & 9.1  & 67& 2  &16.70 & 12.91&0.40& 0.167 &  9.33   &16.20\\
 3385&  4500 & Scd & 13.8 &11.9  &234& 7  &15.74 &  9.86&1.4 & 0.294 & 10.98   &14.75\\
 3468&    --  & Scd & 10.2 & 8.6  &  \multicolumn{2}{c|}{--}     &17.94 & 12.57&0.15& 0.247 &   --     &  --\\
 3515&  6008 & Sd  & 14.3 &10.4  &138&10  &16.89 & 12.27&0.21& 0.225 & 10.26   &  --\\
 3516&    --  & Sc  & 11.1 & 9.7  &  \multicolumn{2}{c|}{--}      &17.79 & 14.07&0.14& 0.169 &   --     &  --\\
 3549&    --  & Sc  & 10.0 & 9.3  &  \multicolumn{2}{c|}{--}      &17.78 & 14.35&0.22& 0.146 &   --     &  --\\
 3558&    --  & Sc  & 12.3 & 9.1  &  \multicolumn{2}{c|}{--}      &18.59 & 13.59&0.10& 0.057 &   --     &  --\\
 3659&  5563 & Scd & 11.2 & 8.7  & 80&3   &17.58 & 14.16&1.13& 0.100 &  9.42   &16.07\\
 3779&    --  & Scd & 11.7 & 8.7  &  \multicolumn{2}{c|}{--}      &18.17 & 13.67&0.10& 0.045 &   --     &  --\\
 3879&  7827 & Scd & 10.6 &10.2  & 79& 3  &17.84 & 15.02&0.16& 0.067 &  9.37   &17.06\\
 4209&  3865 & Sd  & 11.2 &10.1  & 66& 3  &17.29 & 14.06&0.12& 0.089 &  9.15   &16.68\\
 \hline
\end{tabular}
\end{table*}

Dalcanton  and Bernstein \cite{dal2000:Karachentsev_n} 
performed deep photometry of 49 RFGC-galaxies in the $B$-,
\mbox{$R$-,} and $K_s$-bands. Among them there proved to be  25
ultra-flat galaxies, the data on which are given in Table~2. Its
columns contain: (1) the number of the galaxy in the RFGC catalog;
(2)~its heliocentric radial velocity; (3)~morphological type; (4,
5)~apparent axial ratio in the $B$- and $R$-bands from the RFGC;
(6)~the rotation curve amplitude and its error from the HyperLeda;
(7)~the apparent $B$-magnitude within the $27 m_B/\sq\arcsec$
isophote; (8)~the apparent \mbox{$K_s$-magnitude} within the $22
m_K/\sq\arcsec$ isophote and its error; (9)~the $E(B-V)$ color excess
due to Galactic extinction; (10)~the logarithm of the $K$-luminosity
at $m^K_{\odot}=3\fm28$; (11)~the H\,I-flux from the HyperLeda,
expressed in magnitudes.

\begin{figure}
 \onelinecaptionsfalse  \captionstyle{normal} \setcaptionmargin{5mm}
\includegraphics[width=0.85\columnwidth]{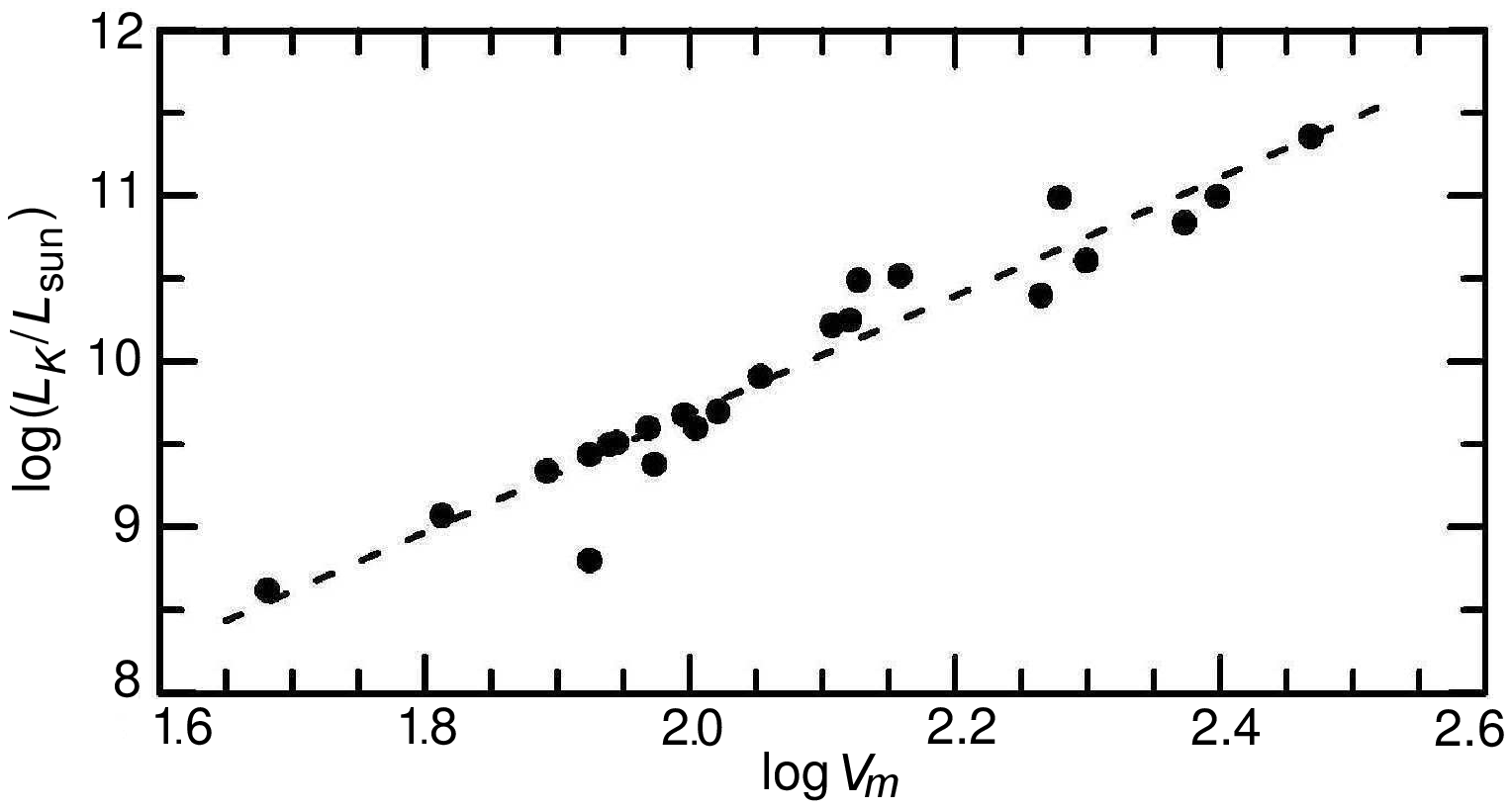}
\includegraphics[width=0.85\columnwidth]{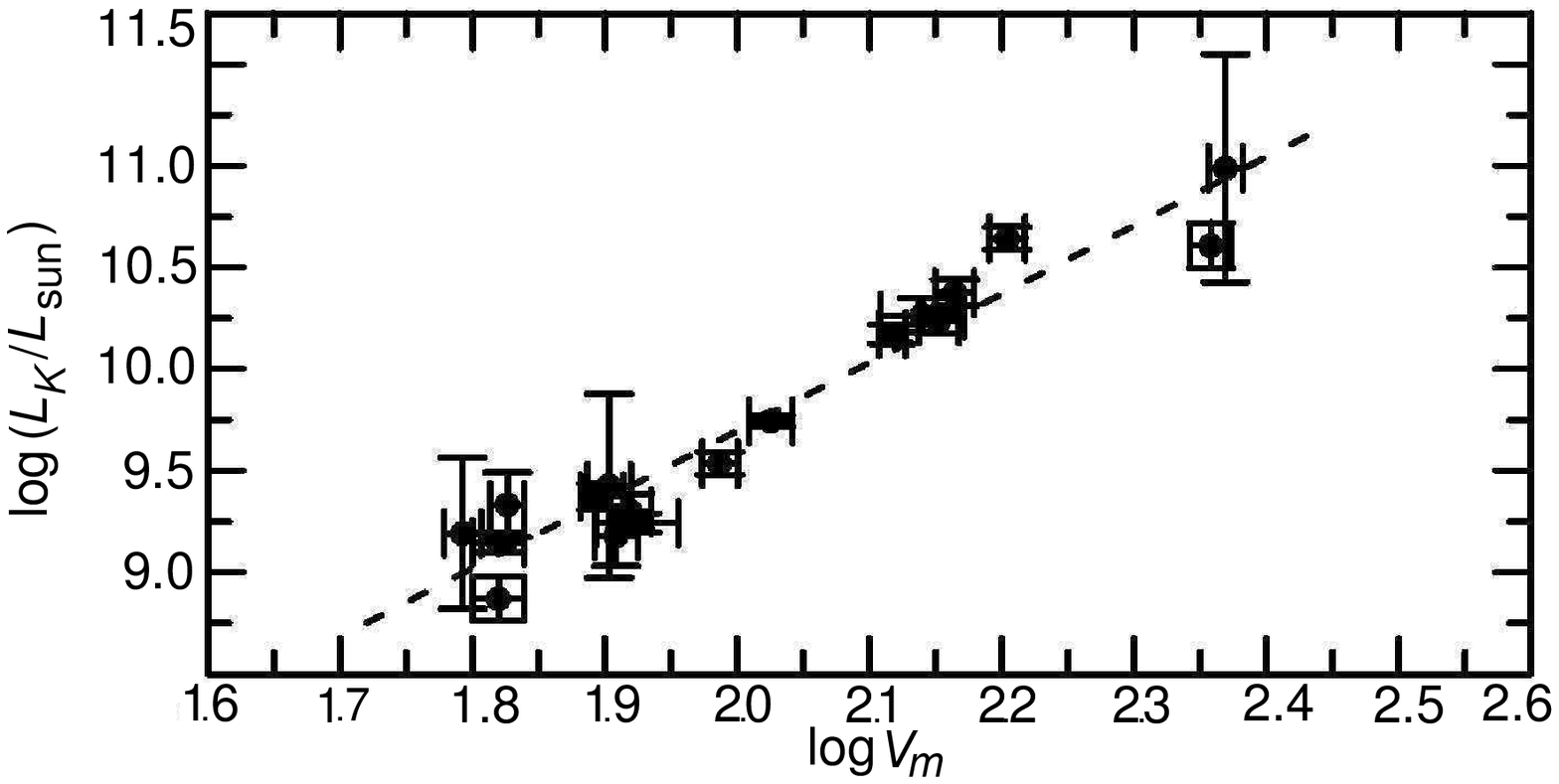}
\includegraphics[width=0.85\columnwidth]{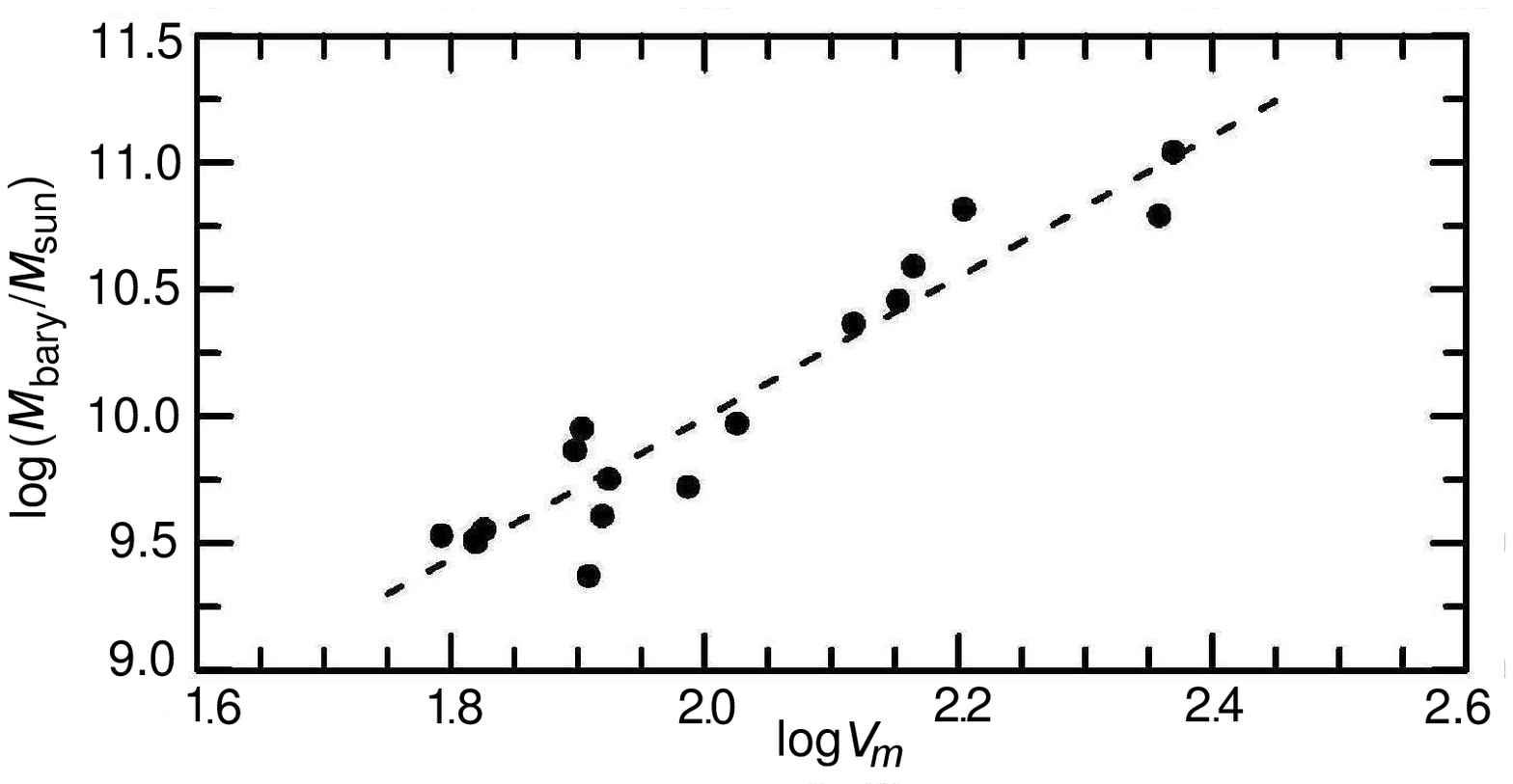}
\caption{The Tully-Fisher relation for ultra-flat galaxies. Top: 23
UF galaxies in Table~1~with the $V_m$ estimates and 2MASS-photometry.
Middle: 18 UF galaxies from Table~2 with an accurate
$K_s$-photometry. Bottom: the baryonic Tully-Fisher relation for  17
UF galaxies with an accurate $K_s$-photometry and  hydrogen mass
estimates. The dashed lines refer to equations (3), (4), and (6) from
top to bottom, respectively.}
\end{figure}

The Tully-Fisher diagram for 18 galaxies with individual
$K_s$-photometry from
\cite{dal2000:Karachentsev_n} 
is presented in the middle panel of Fig.~4. The regression line on it
has the form
\begin{equation}
\begin{aligned}
\log(L_K/L_{\odot})&= 3.37(\pm0.22) \log V_m \\ 
              & +2.96(\pm0.45)
\end{aligned}
\end{equation}
      with the variance of  $\sigma(\log L_K)=0.167$.  The measurement errors of $V_m$ and $L_K$  are shown by the horizontal and vertical bars). The RMS
      errors of the luminosity  and the rotation curve amplitude measurement
for  these galaxies are $\sigma(\log L_K)=0.149$ and $\sigma(\log
V_m)=0.043$, respectively. Given the slope of the regression line of
$3.37$ this gives a total error of $0.207$. Thus, the observed
scatter of galaxies  about the regression~(4) is almost entirely
conditioned by the $L_K$ and $V_m$ measurement errors.
 Taking into account the structural uniformity of UF galaxies, we can expect that their intrinsic (cosmic)
  variance in the  Tully-Fisher diagram is extremely small, and the relation like~(4) is suitable for a reliable determination of   individual distances of UF galaxies.


We have applied relation (4)  to determine the
\mbox{$K$-luminosities} of ultra-flat galaxies in Table~1 via $V_m$
in cases where the 2MASS survey data were not available.

As McGaugh has repeatedly brought
to notice~\cite{mcg2005:Karachentsev_n,mcg2015:Karachentsev_n},
the relationship between  the rotation curve amplitude $V_m$ and the
total baryonic mass of the galaxy, $M_{\rm bary}=M^*+M_{\rm gas}$ has
a more definite physical meaning. Taking into account the correction
for the helium abundance, $M_{\rm gas}=1.33\times M_{\mathrm{H\,I}}$,
and the ratio $M^*/L_K=1M_{\odot}/L_{\odot}$
\cite{bel2003:Karachentsev_n}
        we obtain the relation
\begin{equation}
        \log (M_{\rm gas}/M^*)=0.4 (K_s-m_{21}+2.86), 
\end{equation}
from which it follows that at $m_{21} <K +2\fm86$ the gaseous mass of
a galaxy exceeds its stellar mass. Such objects constitute more than
40\% both in Table~1 and Table~2. The baryonic version of the
Tully-Fisher relation for UF galaxies from Table~2 is shown on the
bottom panel of Fig.~4. The linear regression for them has the form
\begin{equation}
\begin{aligned}
    \log (M_{\rm bary}/M_{\odot})&= 2.78(\pm0.23)\log V_m  \\
                    & +4.44(\pm0.47)  
\end{aligned}
\end{equation}
with the variance  $\sigma(\log M_{\rm bary})=0.172$. A flatter
slope in relation~(6) as compared with (3) and~(4) is caused by a
known fact that dwarf spirals contain much more gas per
  stellar mass unit than disks of high-luminosity galaxies.

\begin{figure}
\onelinecaptionsfalse  \captionstyle{normal} \setcaptionmargin{5mm}
\includegraphics[width=0.9\columnwidth]{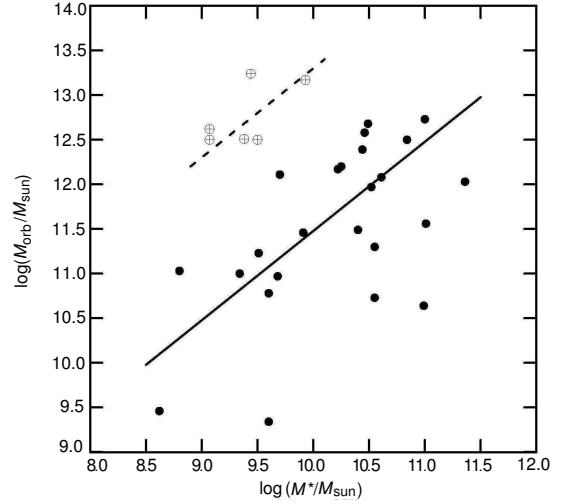}
\caption{The ratio between the stellar mass and the orbital mass
estimate for  UF galaxies from Table~1. The solid line corresponds to
the value of \mbox{$M_{\rm orb}/M^*=30$},  and the dashed line
corresponds to \mbox{$M_{\rm orb}/M^*=2000$} typical for the dark
halo around massive spirals and for associations of dwarf galaxies,
respectively.}
\end{figure}

 Table~3 gives a summary of different average characteristics of UF galaxies
from the samples of Tables~1 and~2 with  errors in mean. As follows
from these data, the samples have approximately the same depth
($V_h$), morphological structure,  apparent axial ratio, rotation
curve amplitude, $K$-luminosity
 and color index ($B-K$), corrected for
Galactic extinction. Therefore, both  samples may well be considered
as taken from a  single general population.

\begin{table}
\setcaptionmargin{0mm} \onelinecaptionstrue \captionstyle{normal}
\caption{The average parameters for two samples of  UF galaxies from
Tables~1 and~2}
\begin{tabular}{l|r@{$\,\pm\,$}l|r@{$\,\pm\,$}l} \hline
  Parameter           &   \multicolumn{2}{c|}{Galaxies}         &  \multicolumn{2}{c}{Galaxies}    \\
                     &    \multicolumn{2}{c|}{Table~2}            &   \multicolumn{2}{c}{Table~1}\\ 
\hline
  $V_h$, km\,s$^{-1}$      &      5100& 620   &   5040& 440\\
  Type               &      6.72& 0.14  &   6.23&0.23\\
 $(a/b)_B$           &     12.77& 0.63  &  11.90&0.28\\
$ (a/b)_R$           &     10.90& 0.51  &  10.52&0.25\\
$ V_{\rm max}$, km\,s$^{-1}$     &       114& 12    &    129&  13\\
$\log(L_K/L_{\odot})$&     9.76 & 0.15  &  10.03& 0.13\\
$(B-K)-E_{(B-K)}$    &     3.33 & 0.16  &   3.54& 0.13\\
$m_{21} - K$         &     3.02 & 0.20  &   3.23& 0.18\\
\hline
\end{tabular}
\end{table}

\begin{figure}
\onelinecaptionsfalse  \captionstyle{normal} \setcaptionmargin{5mm}
\includegraphics[width=0.99\columnwidth]{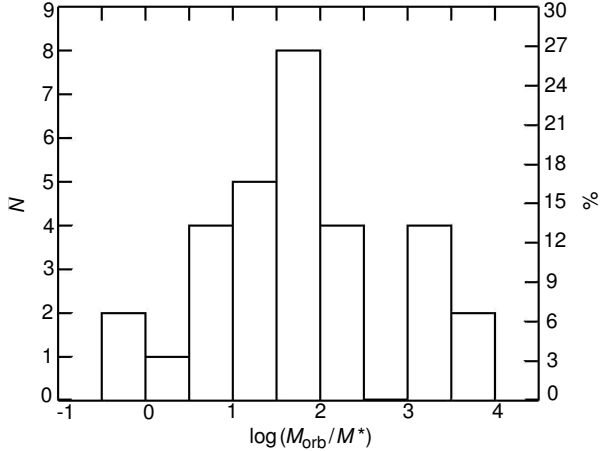}
\caption{The distribution of 30  ``UF galaxy + satellite'' pairs by
the orbital-mass-to-stellar-mass ratio.}
\end{figure}

The distribution of 30  ``UF galaxy + satellite'' pairs by the
orbital and  stellar masses is represented in the
 logarithmic scale in Fig.~5 by circles. The distribution of their number by the $\log(M_{\rm orb}/M^*)$ ratio
  is shown in Fig.~6. In the  range of values of \linebreak $M_{\rm orb}/M^*\sim 10^3-10^4$  there are six pairs,
  consisting of dwarf galaxies, the luminosity of which is fainter than that of the Large Magellanic Cloud.
  According to~\cite{tul2006:Karachentsev_n}
such associations of dwarf galaxies are unbound systems with a formal
virial-mass-to-stellar-mass ratio of approximately  $2\times 10^3$.
The closest example of such a system is a dwarf quartet NGC\,3109 +
Sex\,A + Sex\,B + Antila. In Fig.~5 these six pairs are marked by
light strikethrough symbols. For the other ``UF galaxy + satellite''
pairs there is a distinct tendency to follow the relation
 \mbox{$M_{\rm halo}/M^*\simeq 30$}, found for the brightest spirals in nearby
 groups~\cite{kar+kud2014:Karachentsev_n}.
The scatter in this diagram  is to a large extend due to the
projection effect.

In the presence in the sample of some admixtures of fictitious
unbound pairs,  the most robust estimate of the $M_{\rm orb}/M^*$
ratio is the median estimate, which amounts to $49$ or $32$ depending
on the account or ignoring of 6 suspected unbound pairs.


Here we should note two factors that contribute to the reducing
$M_{\rm orb}/M^*$ estimates. Determining $M_{\rm orb}$, we have
neglected the radial velocity errors. At the median difference in
radial velocities of $87$~km\,s$^{-1}$ and the median error of the
difference  of $34$~km\,s$^{-1}$, an unbiased estimate of $M_{\rm
orb}$ proves to be 15\% smaller than that specified in Table~1.


Calculating the luminosity of ultra-thin galaxies, we have ignored
the correction for the internal extinction in them. According
to~\cite{ver2001:Karachentsev_n}
this correction depends both on the axial ratio of the galaxy and the
rotation curve amplitude:
\begin{equation}
\begin{aligned}
        A_B(int) = (1.54+2.54(\log V_m-2.2))\,\log (a/b). 
\end{aligned}
\end{equation}

At the given in Table~3 mean amplitudes of the rotation curve of
\mbox{$114$--$129$~km\,s$^{-1}$} and the average apparent axial
ratios of \mbox{$10.52$--$12.77$}, the typical internal extinction in
UF galaxies is  \mbox{$A_B(int) = 1\fm22$--$1\fm41$} or $A_K(int) =
0.083 A_B = 0\fm10$--$0\fm12$. Accounting for the internal extinction
would on the average increase the stellar mass by about $11$\%.
Taking into account both corrections,  the median ratio of $M_{\rm
orb}/M^*$ goes down to \mbox{$36$--$24$} in line with the typical
ratio of~$\sim30$ for massive spirals in the nearby groups.


As follows from the data in Table~3, the average morphological type
of UF galaxies on de Vaucouleurs  scale for the two sub-samples is
equal to \mbox{$ 6.5\pm0.22 $}. According
to~\cite{jar2003:Karachentsev_n},
 the average intrinsic color index of   \mbox{$
\langle B-K\rangle_0 = 2.85\pm0.10$} corresponds to  this type.
Comparing this value with the average value of \mbox{$(B-K) -
E(B-K)$}
 in the penultimate line of Table~3, we get
the average observed color excess  in the UF galaxies due to internal
absorption amounting to  \mbox{$ E(B-K)_{\rm int} = 3.44(\pm0.10) -$}
\linebreak $2.85(\pm0.10) = 0.59\pm0.14$ or the average internal
extinction of
 $A_B(int) = E(B-K)/0.917 = \linebreak 0\fm64\pm0\fm15$.
 As we can see, the observed extinction in the disk  of a typical ultra-flat galaxy proves
  to be about 2 times lower than the one of $1\fm22$--$1\fm44$ expected  from relation~(7).
This significant difference may indicate a scarcity of the dust
component in the UF galaxies due to features of their evolution or an
inapplicability of the Verheijen
relation~\cite{ver2001:Karachentsev_n} to very thin disks of
galaxies.

\section{THE NEAREST ULTRA-FLAT GALAXY REPRESENTATIVES}

Considering the cases of the nearest super-thin galaxies, we have the
possibility to estimate in detail the features of  the environment in
which they reside. To this end, we have identified in the
list~\cite{kara2016:Karachentsev_n} four UF galaxies with radial
velocities relative to the centroid of the Local Group amounting to
$V_{\rm LG} < 600$~km\,s$^{-1}$.

{\it RFGC\,1462 $=$ UGC\,4704}. This isolated  Sdm type galaxy with
the radial velocity of \mbox{$V_{\rm LG} = 584$~km\,s$^{-1}$}  and
the apparent magnitude of \mbox{$B=15\fm0$}  has a close dwarf
companion \mbox{($B = 17\fm2$)}   at the projected separation of
$48$~kpc with the velocity difference of \mbox{$\Delta
V=8$~km\,s$^{-1}$}. Apart from it, at a distance of
\mbox{$R_p=574$~kpc} there is another dwarf companion
\mbox{($B=17\fm8$)} with the  radial velocity difference of
$18$~km\,s$^{-1}$, which  we have not included in Table~1 because of
a large projected separation.

{\it RFGC\,1561 $=$ UGC\,5047}. This spiral   Sdm-type galaxy with an
apparent magnitude of $B = 16\fm0$ and radial velocity  of $V_{\rm
LG}=552$~km\,s$^{-1}$  has 17 neighbors within $R_p=750$~kpc in the
range of radial velocities of $\mid \Delta V\mid <500$~km\,s$^{-1}$.
Some of the galaxies in the group are brighter than the UF galaxy
itself, making the estimation of its mass by  the orbital motions of
its neighbors incorrect.

{\it RFGC\,1700 $=$ UGCA\,193}. A dwarf   Sdm-type spiral with the
radial velocity of $V_{\rm LG} = 426$~km\,s$^{-1}$ and the apparent
magnitude of   $B=14\fm7$. It has 16 neighbors in the above range of
$R_p$ and $\mid\Delta V\mid$.  This UF galaxy is a  peripheral
satellite of a massive S0 galaxy NGC\,3115, located at a distance of
\mbox{$D = 9.7$~Mpc.}

{\it RFGC\,2246 $=$ UGC\,7321}. One of the thinnest   Sd-type
galaxies with the radial velocity of \linebreak \mbox{$V_{\rm LG} =
344$~km\,s$^{-1}$} and the apparent magnitude of $B=14\fm1$
thoroughly researched by\linebreak
Matthews~\mbox{\cite{matt1999:Karachentsev_n,matt2000:Karachentsev_n,matt2003a:Karachentsev_n,matt2003b:Karachentsev_n}.}
A subsystem of H\,II regions in it has a record high axial ratio of
\mbox{$(a/b)_{H\,II}=38$} \cite{kar2015:Karachentsev_n}.
The galaxy is located on the far outskirts of the Virgo cluster at a
distance of  \mbox{$D = 17.2$~Mpc} in a  scattered group with other
brighter members~\cite{kar2011:Karachentsev_n}.

{\it RFGC\,2937 $=$ ESO\,274-01}. This isolated Scd galaxy with the
radial velocity of  \linebreak \mbox{$V_{\rm LG} = 337$~km\,s$^{-1}$}
and the apparent magnitude of \mbox{$B=11\fm7$}  is not included in
the list of UF galaxies, since it is located on a low galactic
latitude,  \mbox{$b=9^{\circ}$}. The ratio of the blue diameters of
it is \mbox{$(a/b)_B=10.34$}, but the ratio of red diameters $8.18$,
is a little smaller than the limit of $8.5$ we have
adopted~\cite{kara2016:Karachentsev_n}.
Nevertheless, this  ``almost UF'' galaxy is interesting owing to its
short distance \mbox{$D=3.1\pm0.3$~Mpc}, measured with high accuracy
by the tip of  the red giant branch~\cite{kar2007:Karachentsev_n}. At
the projected separation  of  $160\arcmin$, or $144$~kpc from it a
fainter Im-galaxy ESO\,223-09 is located; the difference in their
radial velocities amounts to \mbox{$\Delta V=64$~km\,s$^{-1}$}.  By
the criterion we have adopted,  this galaxy should be regarded as a
physical companion of RFGC\,2937. However, the distance to
ESO\,223-09 \mbox{$D=6.4\pm0.6$~Mpc}, also  measured by the red giant
branch~\cite{kar2007:Karachentsev_n}, points to an accidental optical
proximity of these galaxies separated by a spatial distance of
\mbox{$3.3\pm0.7$~Mpc}. The presence of such a flat     galaxy in a
sphere with the radius of just $3.1$~Mpc around the Milky Way may be
indicative of a significant abundance of thin disk galaxies in the
local universe.

         \section{FINAL REMARKS}
The study of the properties of edge-on ultra-thin spiral galaxies has
not yet  gained a  required systematic character. Selected UF
galaxies were observed by different authors in the optical and radio
domains~\mbox{\cite{mat2000:Karachentsev_n,huc2005:Karachentsev_n,mak1999:Karachentsev_n,mak2001:Karachentsev_n,dal2000:Karachentsev_n,kre2004:Karachentsev_n}}.
         It has been suggested in~\cite{ban2013:Karachentsev_n} that the extreme flatness of  UF galaxies
          is caused by a specific structure of their dark halo, in particular, a high
              halo-mass-to-stellar-disk-mass ratio.
  However,  the first estimates of the \mbox{$M_{\rm orb}  /M^*$} ratio that we have  presented
for ultra-thin galaxies based on the kinematics of their satellites
do not differ significantly from the typical value of  \mbox{$M_{\rm
halo}/M^*\simeq 30$} for the brightest spirals like M\,31, M\,81 in
the nearby groups. Note that the isolated spiral and elliptical
galaxies from the 2MIG catalog  have approximately the same ratios
\mbox{$M_{\rm orb}  /M^*\sim 17-63$}~\cite{kara2011:Karachentsev_n}.
These galaxies are located in the low density regions, residing often
in diffuse filaments and clouds. To refine
 $M_{\rm orb}  /M^*$ in ultra-thin galaxies, a  systematic search for new dwarf satellites around them with radial velocity
  measurements both in the H\,I line, similar to the  AGES survey~\cite{min2010:Karachentsev_n},
and in the optical spectra is required. As noted above,  the UF
galaxy satellites  are predominantly  dwarf BCD and Im-galaxies, rich
in gas and young stars, which makes them  convenient objects  for
measuring the radial velocity. The list of 817 ultra-flat galaxies we
have presented in~\cite{kara2016:Karachentsev_n}
          is a  good basis for such a program.

\begin{acknowledgements}
In our work we used the  NED and HyperLeda databases. IDK thanks the
Russian Science Foundation  for the support  (grant
No.~14--02--00965).
\end{acknowledgements}


\begin{thebibliography}{99}

\bibitem{kar1989:Karachentsev_n}
I.~{Karachentsev}, \aj \textbf{97}, 1566 (1989).

\bibitem{kau2009:Karachentsev_n}
S.~J.~{Kautsch}, \pasp \textbf{121}, 1297 (2009).

\bibitem{sha2015:Karachentsev_n}
X.~{Shao}, K.~{Disseau}, Y.~B.~{Yang}, et~al., \aaa \textbf{579}, A57
(2015).

\bibitem{mat2000:Karachentsev_n}
L.~D.~{Matthews} and W.~{van Driel}, \aas \textbf{143}, 421 (2000).

\bibitem{kar2002:Karachentsev_n}
I.~D.~{Karachentsev}, S.~N.~{Mitronova}, V.~E.~{Karachentseva},
et~al., \aaa
  \textbf{396}, 431 (2002).

\bibitem{kau+2009:Karachentsev_n}
S.~J.~{Kautsch}, J.~S.~{Gallagher}, and E.~K.~{Grebel}, Astronomische
  Nachrichten \textbf{330}, 1056 (2009).

\bibitem{kre2005:Karachentsev_n}
M.~{Kregel}, P.~C.~{van der Kruit}, and K.~C.~{Freeman}, \mnras
\textbf{358},
  503 (2005).

\bibitem{goa1981:Karachentsev_n}
J.~W.~{Goad} and M.~S.~{Roberts}, \apj \textbf{250}, 79 (1981).

\bibitem{kor2013:Karachentsev_n}
J.~{Kormendy}, \emph{{Secular Evolution in Disk Galaxies}} (2013),
p.~1.

\bibitem{sac2015:Karachentsev_n}
S.~{Sachdeva}, D.~A.~{Gadotti}, K.~{Saha}, and H.~P.~{Singh}, \mnras
  \textbf{451}, 2 (2015).

\bibitem{kara2016:Karachentsev_n}
V.~E.~{Karachentseva}, Y.~N.~{Kudrya}, I.~D.~{Karachentsev}, et~al.,
\ab
  \textbf{71}, 1 (2016).

\bibitem{kar1999:Karachentsev_n}
I.~D.~{Karachentsev}, V.~E.~{Karachentseva}, Y.~N.~{Kudrya}, et~al.,
\bsao \textbf{47} (1999).

\bibitem{kar2014:Karachentsev_n}
I.~D.~{Karachentsev}, E.~I.~{Kaisina}, and D.~I.~{Makarov}, \aj
\textbf{147},
  13 (2014).

\bibitem{kar+kud2015:Karachentsev_n}
I.~D.~{Karachentsev} and Y.~N.~{Kudrya}, Astronomische Naschrichten
  \textbf{336}, 409 (2015).

\bibitem{kar+kud2014:Karachentsev_n}
I.~D.~{Karachentsev} and Y.~N.~{Kudrya}, \aj \textbf{148}, 50 (2014).

\bibitem{jar2000:Karachentsev_n}
T.~H.~{Jarrett}, T.~{Chester}, R.~{Cutri}, et~al., \aj \textbf{119},
2498
  (2000).

\bibitem{jar2003:Karachentsev_n}
T.~H.~{Jarrett}, T.~{Chester}, R.~{Cutri}, et~al., \aj \textbf{125},
525
  (2003).

\bibitem{kar1987:Karachentsev_n}
I.~D.~{Karachentsev}, \emph{{Dvojnye galaktiki (Double galaxies).}}
(1987).

\bibitem{bin1998:Karachentsev_n}
J.~{Binney} and M.~{Merrifield}, \emph{{Galactic Astronomy}} (1998).

\bibitem{sch1998:Karachentsev_n}
D.~J.~{Schlegel}, D.~P.~{Finkbeiner}, and M.~{Davis}, \apj
\textbf{500}, 525
  (1998).

\bibitem{hei1972:Karachentsev_n}
J.~{Heidmann}, N.~{Heidmann}, and G.~{de Vaucouleurs}, Mem. R. Astr.
Soc.
  \textbf{75}, 85 (1972).

\bibitem{kud1994:Karachentsev_n}
Y.~N.~{Kudrya}, I.~D.~{Karachentsev}, V.~E.~{Karachentseva}, and
S.~L.
  {Parnovskii}, Astron.Lett. \textbf{20}, 8 (1994).

\bibitem{kar1997:Karachentsev_n}
Y.~N.~{Kudrya}, V.~E.~{Karachentseva}, and I.~D.~{Karachentsev},
Astron.Lett. \textbf{23}, 633 (1997).

\bibitem{bel2003:Karachentsev_n}
E.~F.~{Bell}, D.~H.~{McIntosh}, N.~{Katz}, and M.~D. {Weinberg},
\apjs
  \textbf{149}, 289 (2003).

\bibitem{tul1977:Karachentsev_n}
R.~B.~{Tully} and J.~R.~{Fisher}, \aaa \textbf{54}, 661 (1977).

\bibitem{dal2000:Karachentsev_n}
J.~J.~{Dalcanton} and R.~A.~{Bernstein}, \aj \textbf{120}, 203
(2000).

\bibitem{mcg2005:Karachentsev_n}
S.~S.~{McGaugh}, \apj \textbf{632}, 859 (2005).

\bibitem{mcg2015:Karachentsev_n}
S.~S.~{McGaugh} and J.~M.~{Schombert}, \apj \textbf{802}, 18 (2015).

\bibitem{tul2006:Karachentsev_n}
R.~B.~{Tully}, L.~{Rizzi}, A.~E.~{Dolphin}, et~al., \aj \textbf{132},
729
  (2006).

\bibitem{ver2001:Karachentsev_n}
M.~A.~W.~{Verheijen}, \apj \textbf{563}, 694 (2001).

\bibitem{matt1999:Karachentsev_n}
L.~D.~Matthews, J.~S.~Gallaher, and W.~van~Driel, \aj {\bf 118},
2751(1999).

\bibitem{matt2000:Karachentsev_n}
L.~D.~Matthews,  \aj {\bf 120}, 1764 (2000).

\bibitem{matt2003a:Karachentsev_n}
J.~M.~Uson and L.~D.~Matthews,  \aj {\bf 125}, 2455 (2003).

\bibitem{matt2003b:Karachentsev_n}
L.~D.~Matthews and K.~Wood, \apj {\bf 593}, 721 (2003).

\bibitem{kar2015:Karachentsev_n}
I.~D.~{Karachentsev}, S.~S.~{Kaisin}, and E.~I.~{Kaisina},
Astrophysics
  \textbf{58}, 453 (2015).

\bibitem{kar2011:Karachentsev_n}
I.~D.~{Karachentsev}, O.~G.~{Nasonova}, and H.~M.~{Courtois}, \apj
  \textbf{743}, 123 (2011).

\bibitem{kar2007:Karachentsev_n}
I.~D.~{Karachentsev}, R.~B.~{Tully}, A.~{Dolphin}, et~al., \aj
\textbf{133},
  504 (2007).

\bibitem{huc2005:Karachentsev_n}
W.~K.~{Huchtmeier}, I.~D.~{Karachentsev}, V.~E.~{Karachentseva},
et~al., \aaa
  \textbf{435}, 459 (2005).

\bibitem{mak1999:Karachentsev_n}
D.~I.~{Makarov}, A.~N.~{Burenkov}, and N.~V.~{Tyurina}, Astron.Lett.
  \textbf{25}, 706 (1999).

\bibitem{mak2001:Karachentsev_n}
D.~I.~{Makarov}, A.~N.~{Burenkov}, and N.~V.~{Tyurina}, Astron.Lett.
  \textbf{27}, 213 (2001).

\bibitem{kre2004:Karachentsev_n}
M.~{Kregel}, P.~C.~{van der Kruit}, and W.~J.~G.~{de Blok}, \mnras
  \textbf{352}, 768 (2004).

\bibitem{ban2013:Karachentsev_n}
A.~{Banerjee} and C.~J.~{Jog}, \mnras \textbf{431}, 582 (2013).

\bibitem{kara2011:Karachentsev_n}
V.~E.~{Karachentseva}, I.~D.~{Karachentsev}, and O.~V.~{Melnyk}, \ab
  \textbf{66}, 389 (2011).

\bibitem{min2010:Karachentsev_n}
R.~F.~{Minchin}, E.~{Momjian}, R.~{Auld}, et~al., \aj \textbf{140},
1093 (2010).

\end{thebibliography}
\end{document}